\documentstyle[floats,twocolumn,epsf,prl,aps]{revtex} 

\draft

\title{Conductivity and Atomic Structure of Isolated Multiwalled Carbon
Nanotubes }

\author{A.Yu. Kasumov$^{1,2}$, H. Bouchiat$^1$, B. Reulet$^1$, O. Stephan$^1$,
I.I. Khodos$^2$,\\ Yu.B.Gorbatov$^2$,  and C. Colliex$^{1,3}$ }

\address{$^1$Laboratoire de Physique des Solides, Associ\'e au CNRS, B\^at 510,
Universit\'e Paris--Sud, 91405, Orsay, France.$ ^2$Institute of
Microelectronics Technology and High Purity Materials, Russian Academy of
Sciences, Chernogolovka 142432 Moscow Region, Russia.$^3$ Laboratoire Aim\'e
Cotton (UPR CNRS 3321), B\^at 505 Universit\'e Paris--Sud,91405, Orsay, France.
\parbox{14cm}{\medskip\rm\small%
  We report associated high resolution transmission electron microscopy  (HRTEM)  and
transport measurements on a series of isolated multiwalled carbon nanotubes. HRTEM observations, by revealing relevant structural features of the tubes, shed some light on the  variety of observed transport behaviors,  from semiconducting  to quasi-metallic type. Non Ohmic behavior is observed for certain samples which exhibit "bamboo like" structural defects. The resistance of the most conducting sample, measured down to 20 mK, exhibits a pronounced maximum at 0.6 K and strong positive magnetoresistance. }}

\begin{document}

\maketitle




\begin{figure}
	\[	\epsfbox{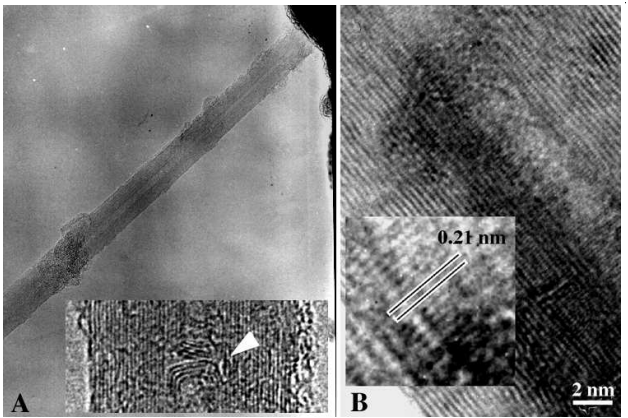}	\]
	\caption{  Transmission electron microscopy on studied samples: A-Partial
view of a connected nanotube (only one metallic pad is visible). Note the presence of some carbonaceous material  visible on the
outer surface of the nanotube. It is mainly due to  electron
damage during observation. Inset: HRTEM of a
bamboo defect.  B-HRTEM picture of Sn2 tube showing evidence of
ordered stacking of non  helical graphite shells.\label{fig1}}
	\end{figure}

Carbon nanotubes discovered by Iijima \cite{iiji} are quite unusual objects
which,  apparently, have no analogs in the solid state. According to
theoretical predictions \cite{mint}, a single graphite layer  wrapped into a
cylinder, can be a metal or a semiconductor depending  on its
diameter and helicity. Due to the peculiarity of the Fermi surface of 2D 
Graphite  which is reduced to a  set of discrete points \cite{wall},  there are
only two conducting channels in a shell independently of its diameter. Thus,
single walled  nanotubes (SWNT), constitute in  principle ideal candidates for the
study of electronic transport at  1D. However they are difficult to isolate 
and manipulate for transport  properties. The experiment by Tans et al.
\cite{tans} showing evidence of single electron  tunneling in an isolated
single-wall nanotube deposited on a silicon substrate is a first fundamental
step in this  direction. However the non metallic character of the electrical
contacts in that  experiment did not allow direct investigation of transport. 
On the other hand multiwalled nanotubes \cite{liu} are easier to manipulate.  
Electron microscopy investigations show that they generally consist of shells
of  different helicity. That is, such a nanotube is a solid body in which one
atomic layer can  be a metal and another  a semiconductor. Transport 
measurements started in 1995,  when some of us  succeeded in measuring the 
conductivity of an individual nanotube \cite{kasu1}. Then, several works on
this  problem were quickly reported \cite{lang,dai,ebe}. In particular Ebessen
et al.\cite{ebe}  have  systematically  studied  the electrical properties of a
large number of nanotubes. It was found that "each  multishell nanotube has
unique conductivity properties". The authors also  suggested that the
difference in electric properties is due to the difference of the  nanotubes
structure. However, even the inner diameter or the number of shells in  these
nanotubes were unknown (this is also pertinent to works described in
\cite{lang,dai}). It seemed  obvious that a detailed investigation of the
nanotubes structure would contribute to the  understanding of 
nanotubes electric properties.

In this letter, we report the results of simultaneous investigations of the
electric  properties and structure of nanotubes in the transmission electron
microscope. The technique  used for isolating  an individual nanotube is
qualitatively  different from the other studies \cite{lang,dai,ebe}. It allows
studying in HRTEM the structure  of the nanotube. The method consists in the
following:  a  focused laser beam "shakes off" a nanotube from the target onto
a sample with a $Si_3N_4$ membrane covered with a metal film \cite{kasu2,kasu3}. A submicron width slit about $100
\mu m$ in length has previously been cut in the membrane by  focused ion beam;
the nanotube connects the edges of the slit see fig.\ref{fig1} and shorts the
electric  circuit whose resistance was over $1 G\Omega$ before the nanotube was
"shaken off".  In the following we successively discuss electron microscopy
observations and transport measurements on a family of tubes indexed as $Au_N$,
$Sn_N$, $Bi_N$ depending on the nature of the used metal contact.  

 The high resolution electron microscope (HRTEM) is a powerful tool to
visualize the atom arrangement in solids. In the case of carbon nanotubes, the
observed contrast can easily be related to the graphitic structure. Graphite
layers ((002) lattice planes), when parallel to the electron beam i.e. in the
Bragg diffraction conditions, are seen as black and white  fringes
corresponding to the projection of atomic positions. A carbon nanotube is
therefore seen as a set of fringes parallel to the tube axis, generated by the
sectors of the coaxial curved graphite sheets which lie parallel to the beam.
Like in graphite, the distance between the layers is approximately 3.4 \AA. We
have used such pictures to estimate the inner and external tube diameters and
the number of layers composing the nanotube. The structural parameters
concerning the whole set of investigated individual tubes are listed in table 1. We have also confirmed that the metal from the
contact does  not wet the internal hollow of the nanotube.

\begin{table}
	\caption { Resistance and structure of various nanotubes.  Note
that the value of the resistance at $100K$ is more adequate to characterize the
transport type of the sample than the value at room temperature.  }
	\label{table} 
	\begin{center} \begin{tabular}{|c|c|c|c|c|c|c|c|} 
	Sample & $\Phi_{in,out}$ & \# of &  L & Bamboo & R $(\Omega)$ &R $(\Omega)$
& Type\\
	name& (nm)  &  shells & ($\mu m$)& defects &$293K $ & 100K & \\ 
	\hline
	$ Au1$ & 25 , 7 &26 &2.1 & No & $1.0\;   10^6$ & $\infty$ & $1_A$ \\
    $Sn2$ & 40 , 3 &55 &0.33 & No & $2.1\;  10^4$ & $\infty$ & $1_A$ \\
    $Sn5$ & 25 , 5 &29 &0.45 & Yes & $2.5\; 10^6 $ & $\infty$ & $1_A$ \\
    \hline
    $ Au2$ & 16 , 5 &16 & 0.36 & Yes & $6.7\; 10^5$ & $10^8$ & $1_B$ \\ 
    $Sn10$ & 7 , 2 &7 &0.35 & Yes & $5.8\; 10^6$ &$ 5\; 10^8$ & $1_B$\\
    $Sn11$ & 13 , 7 &9 &0.17& Yes & $1.8\; 10^6$ &$6\; 10^7 $ & $1_B$\\
    \hline
    $Bi4$& 12 , 4 &11 & 0.4 & Yes & $2.8\; 10^4$&$ 3.4\; 10^4$& 2\\
    $ Au3$ & 25 , 3 &29 & 0.17 & No & $1.9\;  10 ^3$ & $8\; 10^3$ & 2 \\
    $ Au4$ & 26 , 5 &31 & 0.33 & No & $2.0\; 10^5$ &$2.2\; 10^6$ &2 
	\end{tabular}
	\end{center}
	
	 \end{table}

Another structural parameter is the helicity of the layers, i.e. the
way the carbon hexagons paving the graphite sheet are oriented with respect to
the tube axis. In non helical tubes, the hexagons in two sectors diametrically
opposed are parallel. As a result of the hexagonal lattice symmetry, the
contrast in the areas with graphite planes perpendicular to the electron beam
is composed of 3 families of parallel fringes arranged in a three fold symmetry
pattern. The distance between these fringes is 2.1 \AA\  and one family of
fringes is either perpendicular to the tube axis (zig-zag tubes) or parallel to
the tube axis (arm-chair tubes). As an example, we show in fig.1B a high
resolution image of the tube constituting Sn2 sample. In this case, one family
of fringes is more obvious than both others. The fringes are perpendicular to
the tube axis indicating that the cylinders configuration is zig-zag like. This
picture according to simulations by Zhang et al. \cite{zhang} is a strong
indication for a graphite like  stacking of most of the   the nanotube shells. 
 Most investigated nanotubes  cannot be
described only as perfect sets of coaxial cylinders, but exhibit defects which
can affect the transport mechanisms. In particular a defect so called bamboo
defect in the literature \cite{bamboo}, has been identified in many cases: the
inner  shells of the nanotube are  interrupted and  separated by fullerenic
semi-spheres while the outer shells remain continuous, (see fig.1A). The
presence  of such defects is indicated in table 1. We will see that
the presence of these defects affects transport properties when there exists a
possibility of conduction between the outer and inner shells of the tube.

\begin{figure}
	\[	\epsfbox{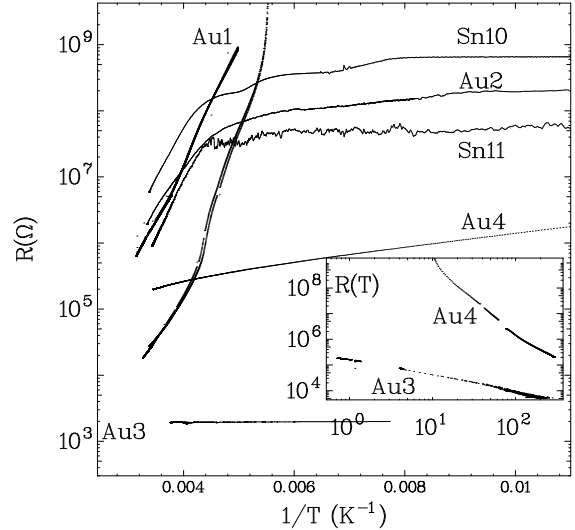}	\]
	\caption{High temperature resistance (on semi-logarithmic scale) of the
tubes, showing  evidence of semiconducting behavior for most of them. Inset: 
Temperature dependence on a Log-Log scale of the resistance of  Au4 and  Au3  
showing evidence of power law increase  at low temperature.  \label{fig2}}
	\end{figure}

\begin{figure}
	\[	\epsfbox{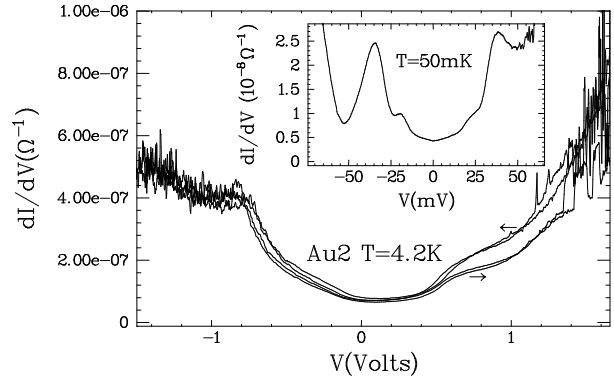}	\]
	\caption{Non linear transport in Au2  on a wide voltage scale  and inset
showing narrow peaks at  low temperature. These measurements were done by
applying a small voltage modulation  superimposed on the dc
voltage.\label{fig3}}
	\end{figure}

 Concerning transport  properties two main types of behavior have been
observed: semi-conducting, (type $1_A$, $1_B$)  and quasi-metallic, (type$2$)
as depicted below:

 Most  samples belong to type $1$. They exhibit a semi-conducting behavior
below $300\;K$,  with exponentially activated temperature dependence  of the
resistance,(see fig.\ref{fig2}). Typical gaps values  lie between 2000 and 3000
K, similar to values obtained in amorphous graphite \cite{robe}.  Note however,
that the room temperature values of the effective resistivities, estimated from
length and section of the tubes  $\rho_{eff}= RS/L$, are all below $1\Omega
cm$ which is orders of magnitude smaller than the typical values obtained for
amorphous graphite.  For some  of these tubes  (type$ 1_B$),   a saturation of
the resistance around $10^8 \Omega$ is observed below 100K.  It is striking  
that all these tubes contain one "bamboo" defect. These samples  are also characterized by their strongly
non linear $I-V$  characteristics  below $100\;K$, see 
fig.\ref{fig3}.  In most cases the $dI/dV$  curves are not symmetrical in $-V,
+V$ and also  exhibit  hysteresis and telegraphic  noise only for one
particular voltage sign.  The typical voltage scale for the non-linear behavior
is of the order of few tenths of volts. At 4.2K and below the differential
conductance exhibits some  narrow peaks characteristic of a Coulomb blockade 
stair-case like behavior   (width of the order of 10mV). For particular  
voltage values, telegraphic noise could also be recorded. The characteristic
time scales are of the order of the ms at 77K and 1000s below 1K.   

\begin{figure}
	\[	\epsfbox{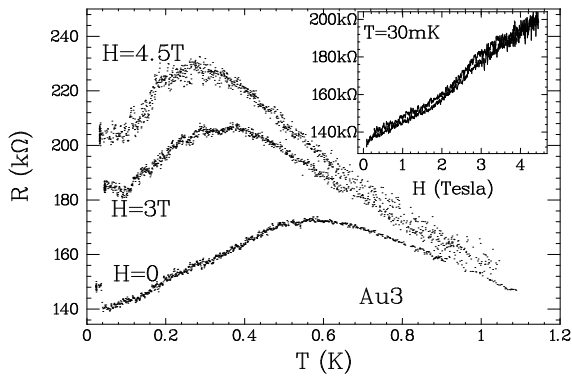}	\]
	\caption{Low temperature resistance on Au3 sample for several values of 
magnetic field. Inset: magnetoresistance at 30mK. The measurements were done
with a ac current of 40pA at 30Hz.  \label{fig4}}
	\end{figure}

  A few tubes  (Au4, Au3 and Bi5) belong to type $2$ and exhibit  a
"quasi-metallic" behavior. Their resistances   increase more slowly than
exponentially at low temperature,  varying   approximatively like $1/T ^x$ (see fig.\ref{fig2} with
$x =0.5$ for Au3 and  $x=2$ for Au4, a lower increase  was observed for Bi5. We
never see any increase of resistance at high  temperature similar to what is
observed in "bulk" samples of SWNT  \cite{fish}. Note however that this type of
true "metallic" behavior has only been recorded  so far in "bulk" samples or
ropes of SWNT and has never been  reported for isolated nanotubes.   On the 
Au3 sample we could also perform very low temperature transport  measurements,
shown in fig.\ref{fig4}. The $R(T)$ curve exhibits a broad maximum around 0.6 
K. The amplitude and position  of this maximum varies drastically with  the
magnitude of magnetic  field applied  perpendicularly to the tube axis. It
shifts  to lower temperature with increasing magnetic field and reaches higher
resistance values. Accordingly, one observes a large positive
magneto-resistance  approximatively linear in magnetic field (with a $ 50\%$ 
increase for an applied field of 4T.) To our knowledge  it is the first time
that such remarkable features have been observed in the resistance 
measurements of nanotubes. These findings are in contrast with those of Langer
et al. \cite{lang} who had measured negative magnetoresistance,  which could be
interpreted as a weak localization effect.   

One important issue for understanding transport properties of these nanotubes
is the separation between the contribution of the most external shell, which is
the only one  directly connected to the  metallic pads, and the possible
contributions of internal shells. This is determined by the ratio
$\eta=R_{ext}/R_t$ between the resistance of this external shell and the
resistance connecting this shell to internal shells.
  The combination of HRTEM observations and transport measurements  suggests
that $\eta << 1$ for type $1_A$ and type $2$  nanotubes. One convincing example
is the Sn2 sample: it contains 60 shells which, according to HRTEM
observations, are arranged for most of them in  a nearly perfect metallic
graphite type of order. However the temperature dependence of its resistance
indicates a semiconducting behavior. These two results can be reconciled if transport  takes place in
the external shell of the tube and if there is no possibility of conduction
through internal metallic shells.
Unfortunately, we do not have any  indication whether semiconducting  behavior
is related to the helical structure of the tube external  shell or to
structural disorder in this shell. 

 The situation is different for type $1_B$ samples, where   figures \ref{fig2}
and \ref{fig3} indicate a residual tunneling conductivity  at  low temperature.
 A possible explanation of these findings could be tunneling on a metallic
inner  shell of the tube, behaving as a Coulomb island,  separated from the
metallic pads by  external shells which are insulating at low temperature but
however offer the possibility of electron transfer through high but finite
tunneling resistance $R_t$.  HRTEM observations  reveal a  complex situation
with the existence of "bamboo" like defects It is more  reasonable to assume
that these nanotubes contain 2 distinct metallic islands and  eventually 3
tunnel junctions. The  capacitance of such junctions of typically $10 nm$ 
dimensions is of the order of $10^{-19}\; F$. On the other hand the capacitance
of a metallic layer of length $100 nm$ and diameter $10\; nm$ is of the order
of $3\;  10^{-17}\;F$. It corresponds to  charging energies of respectively 
$1\;eV$ and $30\;meV$ , i.e.  which are  compatible with the voltage scales of
the features observed in fig.\ref{fig3} on the non linear $I-V$ 
characteristics of these systems. The  existence of telegraphic noise is 
related to the great sensitivity of the conductance of these junctions to
structural  defects which are still mobile at low temperature for specific
values of dc  polarization of the sample. Note also  that the hysteresis in
these  $I-V$ curves observed only for positive values of voltages cannot be of
thermal origin  and may be  due to some  electro-mechanical effect. We finally 
note that bamboo defects are only relevant for these $1_B$ type tubes for which
internal shells play an important role in low temperature transport properties
($\eta>>1$). They also exist in some type $1_A$ and type $2$ samples but we do
not think that they affect their transport properties which are mainly
determined by  the outer shell ($\eta<<1$). 

Concerning type 2 samples, it is interesting to compare our results with the
power-law increase of the  low  temperature resistance predicted  theoretically
in  nanotubes  as a manifestation of electron-electron interactions on  their
transport properties  \cite{bale}. More generally it is the expected behavior
in a  1D conductor whose Fermi liquid behavior is unstable against the effect
of electron-electron interactions \cite{schu}. The power law exponent is not
universal and can be strongly affected by disorder \cite{giam}. The most
striking result of this work  concerns the low temperature anomaly in the
resistance of Au3 and its magnetic field dependence. We cannot exclude a
superconducting fluctuation but the absence of rounding of the anomaly with
magnetic field does not comfort this hypothesis. Another possibility could be a
dimensional crossover as observed in organic conductors \cite{cross}. It would 
occur when  the temperature is of the order of the coupling between the shells.
The magnetic field, by confining back the electrons in a given shell, would
then be responsible of the shift of this crossover to lower temperature. 
  Finally we want to emphasize that contrary to previous studies  on deposited
nanotubes, we are working on suspended structures. This offers the
  possibility of special vibration modes on the  samples (standing waves whose
wave-lengths are determined by the distance between the electrical contacts).
It is noteworthy that the fundamental  mode of  energy $E_0=hv_s/L$ where $v_s$
is the sound velocity along the tube (of the order of  $10^3m/s$) and $L=100nm$
corresponds to a temperature of the order of 1K, close to  the position of the
maximum observed in the resistance of Au3 sample. As  suggested by the work of
Ajiki and Ando \cite{ando},  applying a transverse magnetic  field on a
nanotube is also expected to  produce a lattice distortion and huge positive
magnetoresistance due to the increase of the gap with magnetic field. These
considerations suggest, according to recent theoretical predictions \cite{kane}
a  huge sensitivity of transport properties of nanotubes to mechanical stress
or  distortions and deserve further experimental investigations.

In conclusion  simultaneous HTREM and resistance measurements performed on the
same samples, highlight the importance of internal structural defects when compared to the helicity parameters, in the
mechanism of electron conductivity. We have also shown
that the outer shell determines the resistance for most insulating or
conducting tubes, even if it is not yet possible to investigate specifically
the structure of this shell. Specific "bamboo" type defects could be identified
which are fundamental for the understanding of intermediate  behavior, where
internal shells  contribute to electronic transport through tunnel junctions.
 We have finally demonstrated 
that measuring transport properties of nanotubes is specially interesting at
very low temperatures with the existence of an anomaly in the temperature
dependence, highly  sensitive to the strength of magnetic field.

   We acknowledge our colleagues P. Ajayan, T. Giamarchi, D. Jer\^ome, P.
Launois, R. Moret for help and discussions.  A.K .acknowledges the University
of Orsay for an invited professor position. We have also benefited from a joint
C.N.R.S and Russian Academy of Science program and partial support of NEDO.

\end{document}